\documentclass{aa}  

\usepackage{graphicx}
\usepackage{txfonts}
\usepackage{color}
\usepackage[dvipsnames]{xcolor}
\usepackage{natbib}
\usepackage{bm} 
\usepackage{amssymb} 
\usepackage{mathrsfs} 
\usepackage{ulem}

\newcommand{\be}{\begin{equation}}
\newcommand{\ee}{\end{equation}}
\newcommand{\Medd}{\dot M_{\rm Edd}}

\begin{document}

  \title{The slimming effect of advection on black-hole accretion flows}
   
   \author{J.-P. Lasota
          \inst{1,2}\fnmsep\thanks{\email{lasota@iap.fr}}
          \and
          R. S. S. Vieira
          \inst{2,3}
          \and
  A. S\c{a}dowski
          \inst{4}
\and 
  R. Narayan
        \inst{5}
          \and
  M. A. Abramowicz        
          \inst{2,6}
}
\institute{Institut d'Astrophysique de Paris, CNRS et Sorbonne Universit\'es, UPMC Paris~06, UMR 7095, 98bis Bd Arago, 75014 Paris, France
\and
             Nicolaus Copernicus Astronomical Center, Bartycka 18, 00-716 Warsaw, Poland  
\and
        Instituto de Astronomia, Geof\'{i}sica e Ci\^encias 
Atmosf\'{e}ricas, Universidade de S\~{a}o Paulo, 05508-090, 
S\~{a}o Paulo, SP, Brazil   
         \and
             MIT Kavli Institute for Astrophysics and Space Research,
             77 Massachusetts Ave, Cambridge, MA 02139, USA
         \and
             Harvard-Smithsonian Center for Astrophysics, 60 Garden St., Cambridge, MA 02138, USA          
          \and
             Physics Department, Gothenburg University, SE-412-96 G\"oteborg, Sweden 
          }

   \date{Received ; accepted }
 
  \abstract
  {At super-Eddington rates accretion flows onto black holes have been
    described as slim (aspect ratio $H/R \lesssim 1$) or thick ($H/R
    >1$) discs, also known as tori or (Polish) doughnuts. The relation
    between the two descriptions has never been established, but it was
    commonly believed that at sufficiently high accretion rates slim
    discs inflate, becoming thick.}
  {We wish to establish under what conditions slim accretion flows
    become thick.}
  {We use analytical equations, numerical 1+1 schemes, and numerical
      radiative
      MHD
    codes to describe and compare various accretion flow
    models at very high accretion rates.}
   {We find that the dominant effect of advection at high accretion
     rates precludes slim discs becoming thick. }
   {At super-Eddington rates accretion flows around black holes can
     always be considered slim rather than thick.}

   \keywords{black holes physics -- accretion, accretion discs 
               }

   \maketitle
%

\section{Introduction}
\label{sec:intro}

At high accretion rates discs around compact bodies cease to be
thin\footnote{In this article accretion discs are called \textit{thin} if their aspect
    ratio satisfies $H/R \ll 1$. Discs satisfying $ H/R \lesssim 1$ and $
    H/R > 1$ are called \textit{slim} and \textit{thick}, respectively}. When radiation provides the dominant pressure and opacities
are mainly due to electron scattering, the thin-disc equations imply that the disc thickness increases linearly with the accretion rate $H(R)\propto \dot M$ \citep{SS73,FKR02}, 
where $H(R)$ is the disc semi-thickness at the distance $R$ from the centre. Hence the general belief that with increasing accretion rates discs become very thick:
they are tori rather than discs. Such accretion flows are supposed to be described adequately only by 2D or 3D structures, contrary to thin discs whose properties (including the
observed ones) are  depicted very well by a (1 + 1)D formalism.

However, this conclusion might be not self-consistent because it follows from the thin-disc equations in which $H/R \ll 1$ is assumed and $\mathcal{O}(H/R)$ terms omitted. 
In particular the (proportional to the radial velocity) terms corresponding to advection are neglected. Taking these terms into account, as required for consistency, modifies 
the conclusions about the disc thickness at high accretion rates. This can be clearly seen in the case of advection-dominated accretion flows, known as \textsl{slim discs} in 
the optically thick case, and as \textsl{ADAFs} when accretion flows are optically thin.

The disc thickness depends mainly on temperature through the speed of sound $c_s$, since $H/R\approx c_s/v_{\rm K}$, where $v_{\rm K}$ is 
the Keplerian velocity. Therefore the disc thickness is determined by the efficiency of the cooling mechanism. 
Efficiently cooled optically thick discs are geometrically thin, but for both low and high accretion rates radiative cooling might not be efficient 
enough to ensure the geometrical thinness of the accretion flow. At low accretion rates, optically thin, gas-pressure dominated flows 
cannot be assumed to be geometrically thin; the same is true of high-luminosity (close to Eddington luminosity $L_E=4\pi GMc/\kappa_{\rm es}$, 
where $M$ is the mass of the accreting object and $\kappa_{\rm es}$ the electron-scattering opacity coefficient), 
radiation--pressure dominated accretion configurations when opacity is mainly due to electron scattering. In such cases $H/R \ll 1$ is no 
longer satisfied and the 1+1 scheme not strictly valid. One can, however,  still try to use such a scheme to describe accretion flows as 
long as $H/R \lesssim 1$. In this case one should take into account the $\mathcal{O}(H/R)$ terms that are dropped in the thin-disc approximation. 
This means taking  {\sl advection} of energy and the advective terms in the momentum 
conservation equation into account (which in the thin-disc approximation reduces to $\Omega=\Omega_{\rm K}$, where $\Omega$ is the angular velocity and 
$\Omega_{\rm K}$ its Keplerian form). \citet{Beg79} found that photon-trapping, related to (but not identical to) advection plays an important role 
in the super-Eddington regime and allows (contrary to a wide-spread prejudice) for limitless rates of accretion onto a black hole.

Using a non-$\alpha$ viscosity prescription, \citet{I77} was the first to study optically thin advection-dominated accretion flows, 
later named ADAFs which, in the framework of the $\alpha$-prescription, have been investigated in \citet{ny0,ny1} and \citet{adaf1}. 
In the case of optically thick discs, \citet{PBK81} were the first to solve accretion disc equations including advective terms, but they 
restricted themselves to the $H/R\ll 1$ case. The slim ($H/R \lesssim 1$) disc case has been addressed and solved by \citet{slim1}. 
In all these models a 1+1 scheme was used; only in \citet{olek11,olekthesis,DS11} the vertical and radial slim-disc structures have been coupled 
in a ``pseudo-2D'' approximation. \citep[see also][for a ``complementary'' approach]{Abolmasov15}.

On the other hand there exists a class of accretion flow solutions for
which $H/R > 1$. 
Various versions of these models are known as accretion tori,
Polish doughnuts or fat discs \citep[see][and references therein]{MA05}. Such structures are supposed to correspond to super-Eddington luminosities. 
Indeed, for non-selfgravitating accretion discs the effective (tidal) gravitational force at the surface must be larger than the radiative 
force \citep{MA05}:
\be
\label{eq:gravradforce}
\frac{GMm_p}{R^2}\frac{H}{R}>\frac{\sigma_T}{c}\frac{3GM\dot M}{8\pi R^3}f,
\ee
where  $m_p$ is the proton mass, $\sigma_T$ the Thomson scattering cross-section, and the inner-boundary condition factor $ f=1-\ell_{in}/\ell$, 
where $\ell$ and $\ell_{in}$ are respectively the specific angular momentum and its value at the inner boundary.
From Eq. (\ref{eq:gravradforce}) it is easy to obtain
\be
\label{eq:condhr}
\frac{H}{R}>\frac{3}{4}\frac{1}{\eta}\frac{R_S}{R}\frac{\dot M}{\dot M_{\rm E}}f,
\ee
where $\eta \sim 0.1$ is the gravitational efficiency, $\dot{M}_E={4\pi GM}/{\eta c\kappa_{es}}$ the Eddington accretion rate, 
$\kappa_{es}$ the electron-scattering opacity, and $R_S=2GM/c^2$ the Schwarzschild radius. Clearly, Eq. (\ref{eq:condhr}) 
implies that for super-Eddington accretion rates $\dot M > \dot M_{\rm E}$ accretion flows must be ``fat'',  at least close to the black hole\footnote{For very low accretion rates, \citet{Reesetal82} suggested the existence of optically-thin, radiatively inefficient, ion-supported tori, whose rotational energy is extracted by large-scale magnetic fields.}. 
However, this is true only if the flow is radiatively efficient because in writing the right hand side of Eq. (\ref{eq:gravradforce}) 
one assumes that viscous heating is equal to radiative cooling. In advection-dominated flows this is not the case and the rhs of 
Eq. (\ref{eq:gravradforce}) should be multiplied by the radiative efficiency which, in the case of slim discs or ADAFs, can be 
much lower than one. Since at high accretion rates advection is important, this leaves open the question of how thick realistic accretion 
flows can be or, in other words: can one transform a slim disc into a thick disc by increasing the accretion rate?

The aim of the present article is to answer this question by investigating how well various approximate schemes describe an accretion flow whose geometrical 
thinness is not {\sl a priori} assumed. In Sect. \ref{sec:H/R} we use a simple analytical model to show that in (totally) advection-dominated 
flows the disc height is independent of the accretion rate. The same is true of ADAFs. It is different from the case of the radiation-pressure, 
electron-scattering dominated Shakura-Sunyaev solution in which $H \sim \dot M$. In Sect. \ref{sec:numerical} we show that also for numerical 
slim-disc solutions with coupled vertical and radial structures, $H$ is independent of the accretion rate. We show in Sect. \ref{sec:GRMHD} 
that the recently obtained axisymmetric global General Relativistic Radiation Magnetohydrodynamical (GRRMHD) solutions \citep{olek14}, 
for which no assumption about the value of $H/R$ is made, are also
radiatively inefficient at very high accretion rates and can be 
considered to be slim ($H/R\lesssim 1$) discs.
In Sect. \ref{sec:paczki} we discuss the relations between fat discs ({\sl Polish doughnuts}) and advection dominated flows. 
Sect. \ref{sec:discussion} contains a discussion of the implications of our results and Sect. \ref{sec:summ} our conclusions.

\section{In advection-dominated accretion flows $H/R$ is constant}
\label{sec:H/R}

It is rather straightforward to show (in the 1+1 framework) that in advection-dominated accretion flows (ADAFs and slim discs) 
the aspect ratio $H/R$ at a given distance $R$ must be constant, independent of the accretion rate. For simplicity but without loss of 
generality we assume that the black hole is non-rotating and use the ``pseudo-Newtonian'' approximation of \citet{PW}.

In the case of a non-rotating black hole, the vertically integrated hydrostatic equation leads to \citep{adaf1,L15}
\begin{equation}\label{eq:HR}
   \left(\frac{H}{R}\right)^2= 
\frac{f}{(b^2\Omega_K^2/\Omega)}\frac{c}{R_S\kappa_{es}}\left(\frac{R_S}{R}\right)^2(\alpha\Sigma)^{-1}\left(\frac{\dot{m}}{\eta}\right),
\end{equation}
where $\dot m=\dot{M}/\dot{M}_E$, $\eta=1/16$ is the pseudo-Newtonian efficiency of accretion.
We assume a polytropic relation between the vertical profiles of $p$
and $\rho$.
The disc semi-thickness $z=H$ corresponds to the location of $\rho=p=0$ 
and $b$ is a constant defined through 
$P/\Sigma=b^2H^2\Omega_\perp^2$, where $P$ is the vertically integrated pressure,
$\Sigma$ is the surface density, $\Omega_\perp$ is the vertical epicyclic frequency
(see Eqs. (\ref{eq:vert-eq-PSigma}) and
(\ref{eq:vert-eq-PSigma-hrho})). For a polytropic index N=3 one has b=1/3.  
Hence $h_\rho=bH$, where
\begin{equation}\label{eq:hrh}
   h_\rho\equiv\sqrt{\frac{1}{\Sigma}\int_{-H}^H \rho z^2 dz}
\end{equation}
is the density scale height\footnote{In GRRMHD simulations (see Sect.~\ref{sec:GRMHD}) $h_\rho$ is calculated in spherical coordinates as
\begin{equation*}
\label{hspher}
h_\rho = R \sqrt{\frac{2\pi}{\Sigma}\int_0^{\pi}\rho\vert\theta - \pi/2\vert^2\sqrt{g_{\theta\theta}}\,d\theta}, \ \, \mathrm{and}\, \ \Sigma=2\pi\int_0^{2\pi}\rho\sqrt{g_{\theta\theta}}\,d\theta,
\end{equation*}
where $g_{\theta\theta}$ is the $\theta\theta$ component of the metric tensor $g_{\mu\nu}$. Time averaged data are used to perform the integral.
}. 
The properties of the disc thickness are then determined by the function $\dot m(\Sigma)$ which, when plotted on the $\Sigma-\dot m$ -- plane, 
forms $S$-shaped curves representing disc thermal equilibria.

Let us assume first that the accretion flow is radiation-pressure dominated and that the opacity is solely due to electron scattering. This regime
corresponds to high-rate accretion onto black holes \citep{SS73}.
Using the \citet{PW} pseudo-Newtonian potential and the vertically integrated formalism of \citet{olek11},
the energy conservation equation can be written as
\be
\label{eq:energy}
Q_+=Q_{\rm adv} + Q_-\,.
\ee
In this equation
\begin{equation}
\label{eq:qplus}
     Q_+ = \chi\frac{3}{2}\frac{\kappa_{es}R_S}{c}\left(\frac{R}{R_S}\right)^2\Omega g\left(\frac{\dot{m}}{\eta}\right)
    \end{equation}
is the viscous heating term (per unit surface), with {\mbox{$\chi=\left({cR_S}/{R}\kappa_{es}\right)^2f\Omega$}},  and $g=-{2}/{3}({d\ln\Omega}/{d\ln R})$.

The first rhs term
\begin{equation}
\label{eq:qadv}
   Q_{\rm adv}=\chi \xi \left(\alpha\Sigma\right)^{-1}\left(\frac{\dot{m}}{\eta}\right)^2
\end{equation}
represents advective ``cooling''. The parameter $\xi$ is defined through:
\be
\label{eq:xi}
\xi=\frac{d\ln P}{d\ln R} - 2\frac{d\ln \Sigma}{d\ln R},
\ee
where $P=P_r$ is the vertically integrated total pressure (in our case equal to the integrated radiative pressure $P_r$).

$Q_-$ corresponds to radiative cooling. The form of this term
depends on the disc plasma microphysics.
For an electron-scatterring-opacity dominated, ``optically thick''\footnote{With only scattering opacity present the flow is not 
effectively optically thick.} disc it can be written as\footnote{We are using here the form of this equation from \citet{Katoetal} -- see
their Eq. (3.38), however, for consistency with solutions of the transfer equation the factor ``64'' should be ``16'' \citep[see][]{L15}.}.
\begin{equation}
\label{eq:Qminusthick}
   Q_-=\frac{64\sigma}{3\kappa_{es}}\frac{T_c^4}{\Sigma},
\end{equation}
\citep[][where $T_c$ is the mid-plane temperature]{olekthesis}
which, in the radiation-pressure dominated case, becomes
\begin{equation}
   Q_-=\chi\frac{8 b\kappa_{es}^{1/2}}{I_4}c^{-1/2}R_S^{-3/2}f^{-1/2}R^2\frac{\Omega_K}{\Omega^{1/2}}
   (\alpha\Sigma)^{-1/2}\left(\frac{\dot{m}}{\eta}\right)^{1/2},
\end{equation}
where $I_4= 128/315$  \citep[][]{olekthesis}. The energy equation takes then the form 
 \begin{eqnarray}
\label{eq:sequation}
   & & \xi\left(\frac{\dot{m}}{\eta}\right)^2 - 
   \frac{3}{2}\frac{\kappa_{es}R_s}{c}\left(\frac{R}{R_S}\right)^2\Omega g(\alpha\Sigma)\left(\frac{\dot{m}}{\eta}\right) \\
   &+& \frac{8 b\kappa_{es}^{1/2}}{I_4}c^{-1/2}R_S^{1/2}f^{-1/2}\left(\frac{R}{R_S}\right)^2\frac{\Omega_K}{\Omega^{1/2}}
   (\alpha\Sigma)^{1/2}\left(\frac{\dot{m}}{\eta}\right)^{1/2}=0, \nonumber
  \end{eqnarray}
which is a third order equation for $\dot{m}^{1/2}$ in terms of $\Sigma$.

In the case of one-temperature, bremsstrahlung cooled ADAFs the analogous equation (differing only in the cooling term) is quadratic in 
$\dot m$ \citep[Eq. 5 in][]{adaf1}. \citep[For a simple version of Eqs. \ref{eq:sequation} and the corresponding ADAF equations see][]{L15}.

High and low $\dot m$ solutions of Eq. (\ref{eq:sequation}) correspond to advection-dominated flows:
\begin{itemize}
\item slim discs at high column densities (high optical depths);
\item  ADAFs at low column densities (low optical depths).
\end{itemize}
In the case of ADAFs there exists a maximum allowed rate $\dot m \sim \alpha^2$ above which no optically thin solution exists.
On the other hand there is minimum accretion rate for the slim disc solutions \citep[see e.g.][]{Chenetal95} below which the last
two terms of Eq. (\ref{eq:sequation}) dominate and one obtains the (``inner-region'') Shakura-Sunyaev solution.  For this 
solution $\dot m\sim \Sigma^{-1}$ hence the disc thickness, $H\sim \dot m$, increases with accretion rate.

In advection-dominated solutions $ \dot m\sim \Sigma$ so from Eq. (\ref{eq:HR}) $H/R=$ const., 
independent of the accretion rate. Since $c_s \sim H$ this is also true for 
the speed of sound. \citep[In the self-similar solutions of][this corresponds to the constant $\sqrt{c_3}$.]{ny1} 
{\sl Therefore once they become advection-dominated, accretion flows stop inflating with increasing accretion rate.}

\begin{figure}[t]
  \includegraphics[width=0.96\columnwidth]{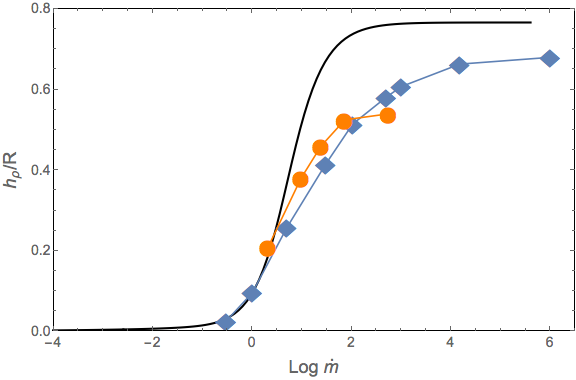} \\
  \caption{Aspect ratio $h_\rho/R$ 
    as a function of accretion rate $\dot{m}$. Solid black line: Calculated at $R=30M$
    for the analytical model described in Section \ref{sec:H/R} assuming
    $M=10M_\odot$, $\xi=1$ and $\alpha=0.01$. Markers: The corresponding aspect ratio for numerical slim disc
    solutions (blue diamonds, Section~\ref{sec:numerical}) and GRRMHD
    simulations (orange circles, Section~\ref{sec:GRMHD}).
  }
  \label{fig:HrhoR} 
 \end{figure}

The black solid line in Figure \ref{fig:HrhoR} shows the relative density scale height $h_\rho/R$ for Keplerian discs in the \citet{PW} potential, for $R=30M$ (in what follows distances are expressed in geometrical units, i.e., with $G=c=1$).
The solution has $\xi=1$, $\alpha=0.01$ and $M=10M_\odot$. 
At the lowest accretion rates the aspect ratio $h_\rho/R$ grows slowly (to the power 1/5) with $\dot m$:  this corresponds to the lower, gas-pressure dominated Shakura-Sunyaev solutions. 
Then $h_\rho/R$ grows almost linearly with $\dot m$ for the (unstable)
radiation-pressure dominated solutions and $h_\rho/R\sim 0.8$ on the slim-disc branch, quickly reaching a saturation value for high $\dot m$.
This behavior does not depend on the particular radius chosen (or on the mass of the central object).

In Fig. \ref{fig:scurvepoints2} the continuous black line represents the solutions of Eq. (\ref{eq:energy}) for accretion discs in which opacity is dominated by 
electron scattering. The two upper branches correspond to the solutions of the Eq. (\ref{eq:sequation}), the uppermost one corresponding to slim-disc solutions.
The lowest branch is gas-pressure dominated and is not described by (\ref{eq:sequation}) but represents the middle-region solution of Shakura \& Sunyaev. 
The lines represent thermal equilibria calculated using the Paczy\'nski-Wiita potential, assuming a Keplerian flow,  $\xi=1$. The black hole mass is $M=10M_\odot$.

Since $h_\rho/R$ depends on $\xi$ and $\Omega/\Omega_K$,  models assuming them to be equal to one could be excessively simple, 
since e.g. as $\dot{m}$ increases and advection starts to dominate (the slim-disc branch) the flow becomes sub- or super Keplerian
depending on the value of $\alpha$ and radius. 
However, using numerical simulations (see below), we checked that these effects do not make much difference and the idealized model gives a pretty accurate description of  the $S$-curves. (see Fig. \ref{fig:scurvepoints2}).

One can conclude therefore that a slim, advection-dominated flow never becomes thick or fat, whatever the accretion rate. The same is true of ADAFs. This is the
simple consequence of the fact that since, in this approximation, \textit{all} the energy gets
advected, not only nothing is left to be radiated away, but also
nothing is left to inflate the disc. The question is how ``realistic'' are such ``totally advection-dominated flows'' (TADAF).
We study this problem in the following sections.

\section{Slim discs}
\label{sec:numerical}

The numerical solutions of stationary (optically thick) slim discs that we use in this work
 were developed in \cite{olek09,olekthesis} as a 
``1+1'' approach to solve the conservation laws equations in general relativity. 
An {\sl a priori} vertical distribution of density and pressure was assumed in order to vertically integrate the disc quantities; 
the resulting ordinary differential equations were then solved numerically in order to obtain the radial dependency of the 
physical parameters of the disc.

Details of the general relativistic model, as well as of numerical techniques, are presented in \cite{olek09,olekthesis}. 
Here we present only the results for a non-rotating BH which are needed for our discussion.

The disc's vertical structure is assumed to obey a relation $p=K\rho^{1+1/N}$, where $p$ is the disc pressure, $\rho$ is its 
rest-mass density, and $N=3$.
The vertical equilibrium equation is then written as
  \begin{equation}\label{eq:vertical-eq-epicyclic}
   \frac{1}{\rho}\frac{\partial p}{\partial z}=-\Omega_\perp^2 z 
  \end{equation}
and gives
  \begin{equation}\label{eq:rho-z}
   \rho=\rho_o\bigg(1-\frac{z^2}{H^2}\bigg)^N,
  \end{equation}
where $z=H$ determines the location of disc surface, and $\Omega_\perp$ is the vertical 
epicyclic frequency of geodesic motion (for zero BH spin, $\Omega_\perp=\Omega_K$).
The temperature profile is assumed to follow \citep{olekthesis}
  \begin{equation}\label{eq:T-z}
   T=T_c\bigg(1-\frac{z^2}{H^2}\bigg).
  \end{equation}
The assumed polytropic ``equation of state'' for the vertical structure implies that 
the disc has a well-defined ``edge'' at $z=H$. At this height (which depends on radius), $\rho=p=T=0$.

Having the vertical profiles, the slim-disc model uses vertically integrated quantities such as $P=\int_{-H}^H p dz$,
$\Sigma=\int_{-H}^H \rho dz$ which satisfy mass, momentum, angular momentum
and energy conservation laws. Vertical integration of Eq.~(\ref{eq:vertical-eq-epicyclic}) implies that
$P$ and $\Sigma$ obey the vertical hydrostatic equilibrium equation
  \begin{equation}\label{eq:vert-eq-PSigma}
   \frac{P}{\Sigma}=b^2 \Omega_\perp^2 H^2,
  \end{equation}
with $b^2=1/(2N+3)$ \citep{olekthesis}.

The scale height of the disc is, of course, smaller than $H$. A good definition of a 
scale height,
taking into account the vertical distribution of matter and/or 
pressure, should give a reasonable measure of the disc thickness. We consider the
density scale height $h_\rho$ (Eq. \ref{eq:hrh})
as a good estimate of the scale height, since it takes into account the variance of the density distribution. It can be shown 
that, for the above vertical profiles, $h_\rho=b H$ (independent of N), and then
  \begin{equation}\label{eq:vert-eq-PSigma-hrho}
   \frac{P}{\Sigma}=\Omega_\perp^2 h_\rho^2.
  \end{equation}
\begin{figure}[t]
 \includegraphics[scale=0.7]{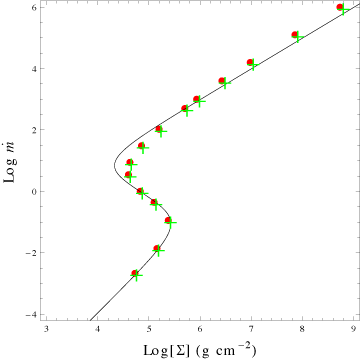}
 \caption{ Thermal equilibrium ($\dot{m}$--$\Sigma$) diagram for slim-disc solutions at $R=30M$. 
 Red dots: Numerical solutions from the slim-disc code \citep{olekthesis}. Black line: Analytical $S$-curve, obtained from the 
 Paczy\'nski-Wiita potential. Both sets have 
 $M=10M_\odot$ and $\alpha=0.01$. The analytical $S$-curve has also $\xi=1$, $\Omega/\Omega_K=1$.
 Green crosses: obtained from the Paczy\'nski-Wiita potential but with the same $\xi$, $\Omega/\Omega_K$ values as the 
 corresponding slim-disc simulations.
 Although the black line does deviate from the red points, we see that when the above corrections are taken into account, the 
 analytical pseudo-Newtonian slim discs agree with the stationary, general relativistic, ``1+1'' simulations with great precision.
 }
 \label{fig:scurvepoints2}
\end{figure}

The general form of the conservation laws (in Kerr geometry) is described in \cite{olekthesis}. For a Schwarzschild black hole we find, as expected, 
that local properties are well described by the Paczy\'nski-Wiita potential in regions not too close to the black hole. 
In particular, the $\dot m$--$\Sigma$ relation was recovered for different values of the radial coordinate. 
Figure \ref{fig:scurvepoints2}
shows one of these $S$-curves, for a Schwarzschild black hole of mass $M=10M_\odot$ and for a radius $R=30M$. The red dots correspond to 
numerical slim-disc solutions constructed for $\alpha=0.01$ and clearly form an $S$-curve, ranging from the lower (stable) gas-pressure dominated 
Shakura-Sunyaev branch 
to the (middle) Shakura-Sunyaev radiation-pressure dominated solutions and the upper slim-disc branch. 
Comparison with analytical Keplerian pseudo-Newtonian $S$-curves 
(with $\xi=1$, black line in Fig. \ref{fig:scurvepoints2}) shows reasonable agreement. 

If one takes 
into account the non-Keplerianity of orbits and the correct values of $\xi$ given by the simulations, calculated via
  \begin{equation}
   Q_{\rm adv}=\frac{- v_R \Sigma}{R}\frac{P}{\Sigma}\xi, 
  \end{equation}
one obtains solutions represented by green crosses in Fig. \ref{fig:scurvepoints2}.
The agreement is then almost perfect. 
Therefore the differences between the two approaches are not due directly to relativistic corrections, 
but result from the non-Keplerian character of the flow and the different values of $\xi$ for each solution. Similar results were obtained for $R=100M$.

As in the case of analytical solutions the slim-disc numerical model \citep{slim1} predicts that the value of the relative disc scale-height
$h_{\rho}/R$ will eventually saturate with increasing $\dot{m}$. 
Moreover, in the pseudo-Newtonian approximation, the behavior of $h_\rho/R$ as a function of $R$ is predicted to follow
  \begin{equation}\label{eq:aspectratioADV}
   \bigg(\frac{h_\rho}{R}\bigg)^2=\frac{3}{2}\frac{fg}{\xi(\Omega_K/\Omega)^2},
  \end{equation}
obtained from Eqs. (\ref{eq:HR}) and (\ref{eq:energy}) in the advection-dominated branch. This expression would be exactly constant
for $\xi=1$ and $\Omega=\Omega_K$. 
Figure  \ref{fig:HrhoR} shows the behavior of $h_\rho/R$
in numerical solutions of slim discs, 
for different values of $\dot m$. As expected, the disc aspect ratio
saturates for the highest accretion rates.
For the chosen value of $\alpha=0.01$, this plateau has $h_\rho/R\approx 0.7$, thus characterizing a slim disc.
The 1+1 numerical scheme preserves the ``no inflation with accretion rate property'' of slim discs.

Since we use the density scale height to determine the disc thickness, one should be sure that the vertical hydrostatic equilibrium equation used,
is an adequate representation of the action of the relativistic tidal ``force'' on the disc's gas.

\citet{Gu07} raised doubts about the validity of the vertical equilibrium equation used in the slim
disc approach but their worries had been already answered in details  by \citet{ALP}. There is nothing wrong with the slim-disc vertical equilibrium equation in
spherical coordinates. Cylindrical coordinates, however, introduce artificial singularities, noticed by \citet{Gu07}, 
but all slim disc models in the Kerr geometry have been calculated in the spherical coordinates in which the
problems discussed in their paper are absent.

\section{Global Radiation - MHD Models}
\label{sec:GRMHD}
The methods discussed so far make 
significant assumptions about disc vertical structure and viscosity.
They separate the radial and vertical dimensions. Only for very thin discs this is
not constraining. To solve geometrically thick accretion flows consistently, one has
to perform numerical simulations in at least 2D. It is a relatively simple
task for optically thin ADAFs, for which the evolution of gas
is independent of the radiative field. In case of optically thick
(super-Eddington) discs, however, the radiation field must be evolved
in parallel to the gas, as it significantly affects its dynamics.
Initial, pioneering work \citep[e.g.][]{Ohsuga+05} used the Newtonian approximation but recently, numerical methods for handling
radiation magnetohydrodynamics (RMHD) in general relativity
have been developed and applied to global simulations of accretion
discs \citep{fragile+14,mckinney+harmrad,olek14}. In this work we use solutions
for non-rotating BHs presented in \cite{olek14}, who performed axisymmetric, long duration
simulations using a mean-field dynamo model to maintain turbulent magnetic
field\footnote{These solutions are in qualitative agreement with
most recent, fully three-dimensional models \citep{sadowski+3d}.}. For a detailed discussion of these simulations we refer the
reader to the original paper. Below we only briefly summarize their
properties. 

The simulations were initiated as equilibrium tori threaded by
multiple loops of a weak poloidal field. The Magnetorotational
Instability (MRI) quickly breaks the equilibrium and makes the gas and
magnetic field
turbulent. This leads to reconnections heating the gas, which in turn
cools itself emitting photons. Because the simulations were
axisymmetric (which allowed for high resolution and long duration), a
mean-field model of the magnetic dynamo has been incorporated to prevent 
the magnetic field from decaying. In this model the poloidal and toroidal
components are modified (conserving energy and momentum, and
preserving
divergence-free condition) to drive the magnetic field towards the
prescribed configuration described by the mean magnetic field angle 
$\theta\approx b^{\hat r}/b^{\hat \varphi}=0.25$ and the magnetic to
gas pressure ratio $\beta'=0.1$. It can be shown that the product of
these quantities determines the order of magnitude of the viscosity parameter
$\alpha=T^{\hat r \hat \varphi}/P\approx \theta \beta'=0.025$. The
actual values of that parameter at radius $R=30M$ are given in the
fourth column of Table~\ref{tab:alphas}.

\begin{figure}[h!]
  \includegraphics[scale=0.25]{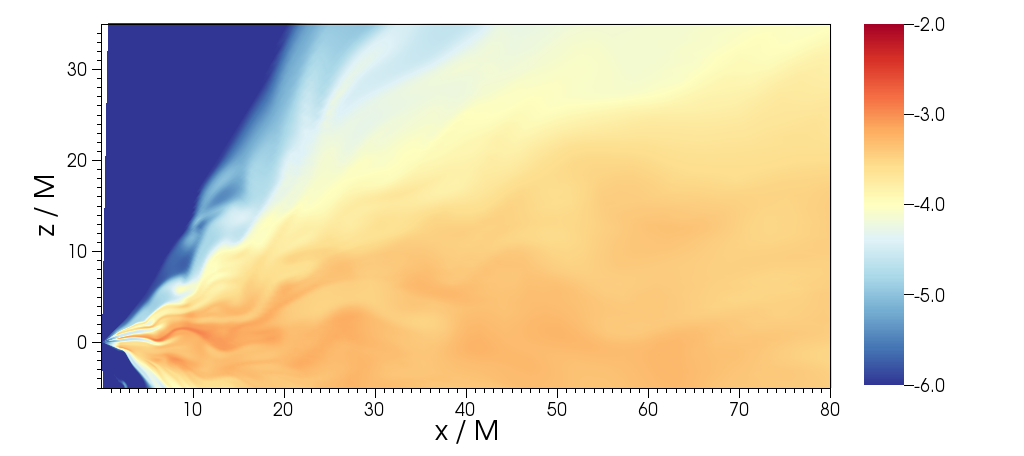}
 \includegraphics[scale=0.25]{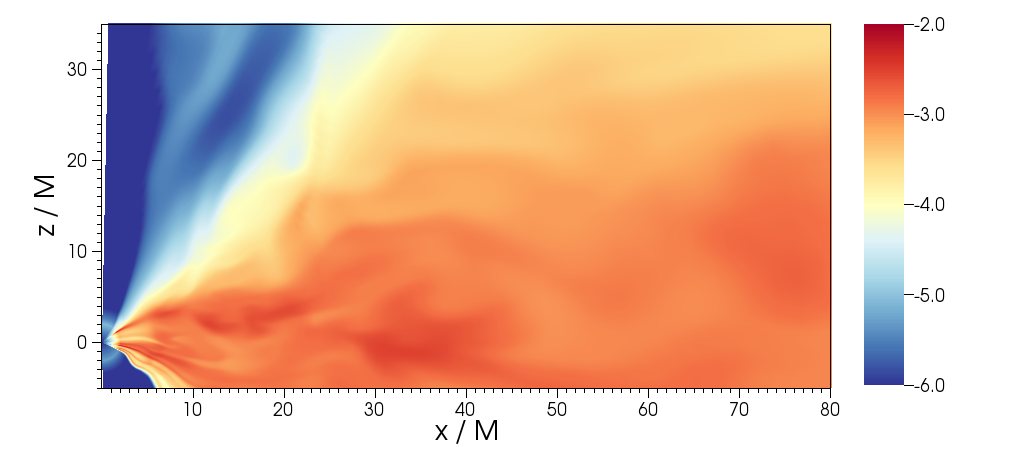}
 \includegraphics[scale=0.25]{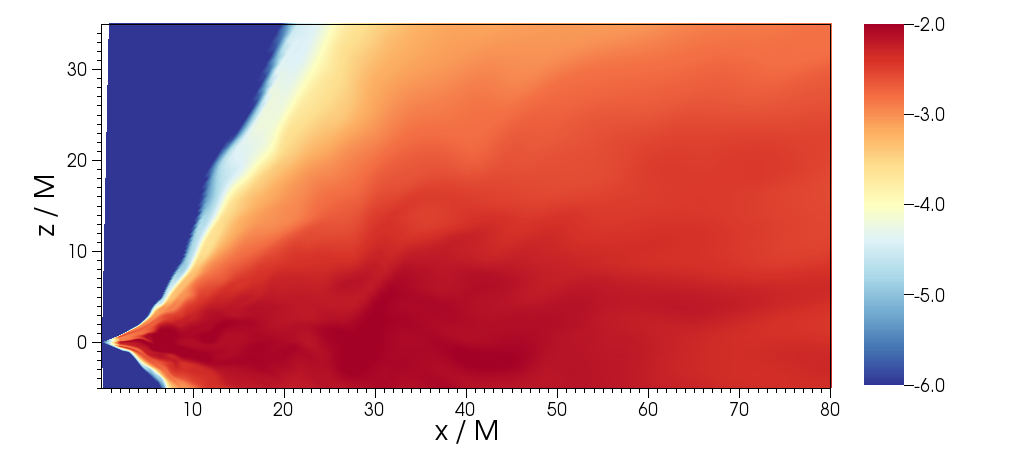}
 \caption{ Snapshots of logarithm of density for  
 GRRMHD discs with $\dot{m}=9.6$, $24.3$, and $73.1\Medd$. Density is
 given in $g/cm^3$.}
 \label{fig:r299a0_hr}
\end{figure}

For our purposes, we consider five models from \cite{olek14}
that are listed in Table~\ref{tab:alphas}. They all exceed the
critical accretion rate and span between $2.1\Medd$ and $558.8\Medd$.
We have chosen as representative $R=30M$ because it is well inside the 
equilibrium region (outflows can be neglected) and far enough from the horizon so that the effects of
disc's puffing-up by magnetic support can be neglected (these magnetic effects cannot be
represented by the slim-disc approximation).
Fig.~\ref{fig:r299a0_hr} presents snapshots of the density for the
  three models with intermediate accretion rates (from top to bottom: 
  $9.6$, $24.3$, and $73.1\Medd$). It is clear that higher
  accretion rates imply higher densities of the gas. The disc
  density scale height, however, does not increase
  significantly, despite the order of magnitude difference in
  accretion rates. This fact is discussed quantitatively below. 

We compare those results now with the properties of
stationary slim-disc simulations (Section~\ref{sec:numerical}).  These comparisons are done in two
ways: by comparing the disc thickness (described by the $h_\rho/R$
ratio), and by comparing locations of the solutions
on the $\dot
m$--$\Sigma$ diagram.
These comparisons must
be qualitative, since the effective value of $\alpha$ varies between
GRRMHD solutions (see Table \ref{tab:alphas}) and is only of the
  same order of magnitude as $\alpha$ assumed for the slim discs
  ($\alpha=0.01$).

Figure~\ref{fig:HrhoR} shows the density scale
  height for slim disc (blue diamonds) and GRRMHD solutions (orange circles). 
Their respective values are very close and in both cases the disc thickness saturates with accretion rate, i.e.,
going to highest accretion rate does not make both discs thicker. The
difference in the actual values comes from the fact that the slim disc
model assumes the vertical structure {\sl a priori}, which does not exactly
reflect the outcome of the turbulent discs. The disc thickness saturation effect is generic,
independent of $R$.

\begin{figure}[ht]
 \includegraphics[scale=0.7]{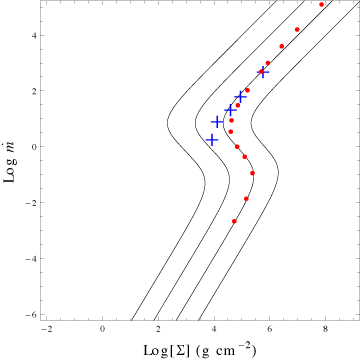}
  \caption{ Thermal equilibrium ($\dot{m}$--$\Sigma$) diagram for $R=30M$. 
  Solutions based on different assumptions are compared. Black lines show analytical $S$-curves for Keplerian motion
  in the Paczy\'nski-Wiita potential with $\xi=1$ and 
  $\alpha=1;0.1;0.01;0.001$ (left to right).
  Red dots represent stationary numerical slim discs for different accretion rates, with $\alpha=0.01$ (Sect. \ref{sec:numerical}). 
  Blue crosses correspond to GRRMHD simulations (see Table \ref{tab:alphas}
  and Sect. \ref{sec:GRMHD}). 
 All solutions have $M=10M_\odot$.
  }
 \label{fig:koralpts30M}
\end{figure}

Another way to compare these two models is by means of a local $\dot
m$--$\Sigma$ diagram shown in 
Fig. \ref{fig:koralpts30M} corresponding to $R=30M$,  zero BH spin,
and $M=10M_\odot$. The previously obtained analytical $S$-curves (obtained from the
Paczy\'nski-Wiita potential assuming Keplerian angular velocity and
$\xi=1$ ) are
shown with
solid lines, and numerical
slim disc solutions with red points. The GRRMHD solutions discussed in
this section are denoted by blue
crosses and show proportionality between the accretion rate and local surface density, 
which is a characteristic
feature of the top, advection-dominated branch (see Sect. \ref{sec:H/R}). What is more,
these points lie almost on top of the slim disc solutions (red dots).
This (almost) exact correspondence is, to some extent, a coincidence because 
the effective viscosity parameters in the two-dimensional, GRRMHD
simulations are not exactly equal to the value assumed in the slim disc solutions ($\alpha=0.01$).

However, since $\alpha$ does not vary much (it is consistent to a
factor of $4$), it is fair to say that GRRMHD solutions belong to the
same advection-dominated, slim disc branch. This is confirmed by the fact that for both
models, the disc thickness saturates with accretion rate.

This agreement between the turbulent discs and the simplified
  model of slim discs is not unexpected. Numerical solutions show
  significant photon trapping inside the disc, which effectively cools
  the disc by advection. They also show outflows which, for
  non-rotating BHs, start only from outside radii $R\gtrsim 30$
  \citep{olek14,SLAN}, and which are not accounted for in the slim disc
  model. Similarly to photon trapping, one may consider the
  outflowing gas as another way of cooling the disc. Both factors make
  numerical solutions radiatively inefficient.
  
It should be stressed that the relation between the hydrostatic shape of the disc and the transfer of radiation
is not straightforward. 
In all GRRMHD models, $h_{\rho}$ is well inside the disc's photosphere,  but 
for $\dot m=24.3$ the base of the optically thin funnel is already far from the black hole at $z=1000M$. 
For accretion rates  $\dot m=73.1$ and $\dot m=558.8$, the whole domain out to $R_{\rm max}= 5000M$ is optically thick at all
$\theta$, so that these two models have no optically thin funnel within the simulation box \citep[][]{olek14}\footnote{\cite[see also][]{sadowski+radjets}.}.
 
\begin{table}
\caption{GRRMHD models \citep{olek14}}
\hspace{.5cm}
\begin{tabular}{ l c c c c }
\hline
  Name &$\dot{m}$ & $\Sigma$ (g cm$^{-2}$)  & $\alpha$ \\\hline \hline
  r299a0 & 2.1 & 8600 &0.022 \\ 
  r300a0 & 9.6 & 13200 & 0.047 \\ 
  r301a0 & 24.3 & 39400 & 0.0515 \\
  r3015a0 & 73.1 & 90000 &  0.0597 \\
  r302a0 & 558.8 & 577300 & 0.081 \\ 
\hline
\multicolumn{4}{l}{$\Sigma$ and $\alpha$ given at $R=30$.}
\end{tabular}
 \label{tab:alphas}
\end{table}

\section{Polish doughnuts}
\label{sec:paczki}

{\sl Polish doughnut} (hereafer PD),
also known as ``thick accretion disc'', is a term describing a model of  axisymmetric, stationary accretion
structure around a black hole, developed in the early days of
accretion astrophysics by the Warsaw group around Bohdan Paczy\'nski
\citep{AJS,PW,JAP} {\cite[see also][]{fishbone}}\footnote{For an excellent pedagogical explanation of PDs see \citet{FKR02}, 
for a review of relevant recent papers \citet{AF}, or \citet{RZ13}. \citet{K06} constructed models of magnetized PDs.}. 
PDs are defined basically by the following three properties:
\begin{enumerate}

\item These optically thick structures radiate locally at the
  Eddington rate, i.e., the local effective gravity at the photosphere
  is balanced by the 
  radiation pressure exerted by 
radiative flux from the photosphere.\\

\item {The shape $H(R)$ of a PD 
resembles a huge spheroid} with long and narrow funnels along the rotation axis. 
Narrow funnels can collimate radiation to super-Eddington
luminosities ($\lambda \equiv L/L_{\rm Edd} > 10$), as was realized
long ago by \citet{S81} and \citet{AP80} \citep[see][for numerical
verification]{sadowski+radjets}.
They are presently of interest for modelling ULXs, \citep[ultra-luminous X-ray sources, e.g.][]{K09,KL14,LKD15}. In such funnels, the relative vertical thickness of a PD 
must be necessarily large, $\chi\equiv (H/R)_{\rm max} \gg 1$ and large PD's thickness is thus unavoidable.\\

\item All the observable properties of a PD are derived from a single
  {\it ad hoc assumed} function, $\ell
= \ell(R)$: the specific angular momentum distribution at the PD's photosphere. The specific angular momentum is assumed to be Keplerian, $\ell(R_{\rm in}) = \ell_K(R_{\rm in})$, at the inner PD edge $R_{\rm in}$ located
between the ISCO and IBCO\footnote{Innermost bound circular orbit.},  which
implies a saddle point in the equipressure surfaces there (a {\sl cusp}). This forces the angular momentum to be Keplerian
also at the PD's ``centre'' corresponding to the {\it pressure maximum}. From the assumed $\ell(R)$ one calculates the PD shape
$H(R)$, the flux of radiation at the surface $\vec{f}(R)$, the total luminosity $L = \int \vec{f} \cdot d\vec{S}$ and finally the
accretion rate is deduced from the luminosity: ${\dot M} = L/(c^2 \epsilon)$, where $\epsilon = \epsilon(R_{\rm in})$ is the radiative accretion efficiency. All these are
given in terms of analytic (algebraic) formulae. {\sl No physical properties of the PD interior need to be considered.} Indeed,
one must stress that the PD models do not assume anything specific about their interiors, not even about the
equation of state, $p = p(\rho, T)$. In particular, the pressure (gas and radiation) $p$, the density $\rho$, the temperature $T$
and the (non-azimuthal) velocity $\vec{v}$ do not appear in the model.
\end{enumerate}

In the PD models it was assumed  that  $H(R)$ corresponds to the photosphere location, however, as checked by \citet{AHG83}, this in general is not the case.
Therefore, as in the case of GRRMHD simulations discussed above, but for different reasons, the relation between the shape of the accretion flow and the radiation transfer is not straightforward. 

The PD formalism has the advantage of allowing constructing thick disc models without having to deal with the largely unknown (especially when they were devised) physics of their interiors. This advantage turns into a drawback when physical processes in accretion flows are better understood and numerical models representing structures analogous to PDs become available. For example, the strength of advective cooling in a PD cannot be calculated from properties 1, 2, and 3 above and no previously calculated analytic models of PDs contained advection. Although numerical models reproduce many properties of analytic PDs \citep[e.g.][]{DVH,Qianetal09}, as noticed by \citet{AF} pressure gradients in PDs are shallower than those in the corresponding numerical simulations. This should mean that numerical models including advection are {\sl thinner} than analytical PD models with no advection.
To test this conclusion \citet{Wetal15} generalized the PD formalism by including an advection-cooling parameter 
\begin{equation}
\zeta = \frac{L_{\rm adv}}{L_0} = \frac{\rm advective~energy~losses}{\rm total~energy~generation~rate}, \ \ \ \ 
0 \le \zeta \le 1.
\label{advective-cooling-parameter}
\end{equation}
The accretion rate is then obtained from
\begin{equation}
{\dot M} = \frac{1}{c^2\epsilon(1 - \zeta)} L.
\label{accretion-rate}
\end{equation}
\begin{figure}[h]
\includegraphics[width=0.45\textwidth]{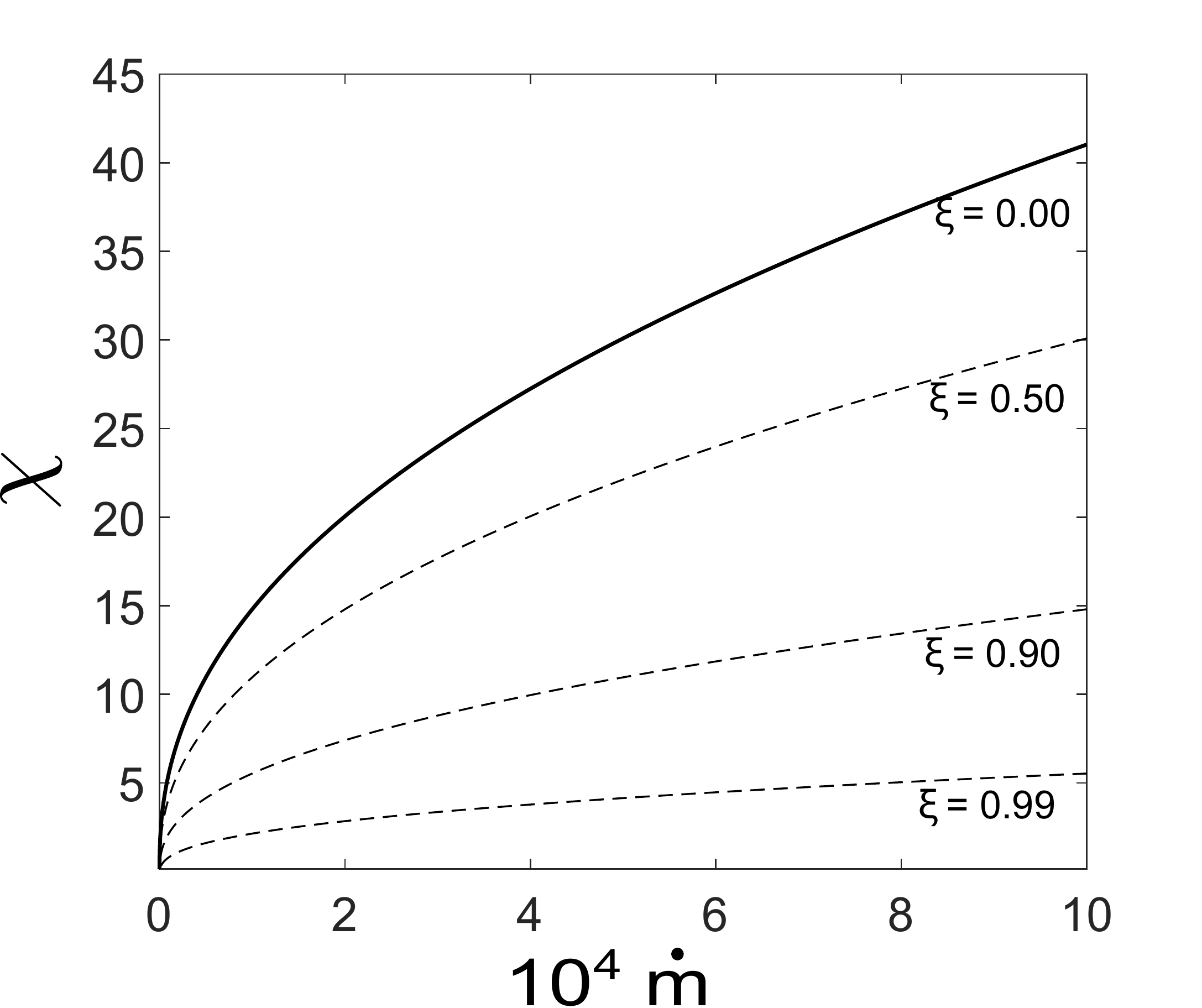}
 \label{fig:Wielgus}
\caption{{Effect of advection on the thickness of PDs. The maximum
  value of the relative height of a PD and the accretion
  rate are shown on the
  vertical and horizontal axes, respectively.} \citep[From][]{Wetal15}.
}
\end{figure}
The relative thickness of these generalized, advective PDs is plotted in Fig.~\ref{fig:Wielgus} which shows
the slimming effects of advection on Polish doughnuts.

A toy model by \citet{BP} illustrates the properties of an extreme PD. In this model \textit{no} energy is
advected because all of it is used to inflate the disc through the $p{\rm d}V$ work \citep[see also][]{adiosM}\footnote{This also pushes the inner flow radius towards the radius of the IBCO.}. 
Since the $p{\rm d}V$ work appears in the advective term of the energy conservation equation, Paczy\'nski called his solution ``advection-dominated'' because
radiative cooling was put strictly equal to zero, but obviously advective-cooling is also equal to zero ($\zeta=0$). There is also no outflow of matter. Although of great heuristic interest, it is unlikely that  Paczy\'nski's toy-model can represent any realistic accretion flow since it requires the angular momentum to be almost constant and strongly sub-Keplerian everywhere except near the outer edge,
where it rapidly joins the local Keplerian value. For example, numerical simulations show that whenever efficient MRI viscosity appears in a constant angular-momentum accretion flow, the angular momentum very quickly relaxes to Keplerian \citep[see, e.g.,][]{Hawley2000}.

A new version of PDs was recently proposed by \citet{ZEBRA} who considered low angular--momentum super-Eddington accretion flow appearing during so-called tidal disruption events 
(TDEs), when stars are captured by a supermassive black hole in a galactic centre.  As in the case of Paczy\'nski's toy model, the accretion energy inflates the flow into a weakly 
bound, quasi-spherical structure. When the flow is maximally inflated, it escapes in form of powerful jets. Also in 
these ZEBRA (zero Bernoulli accretion) flows the angular momentum distribution must  have a specific non-Keplerian form. It still not clear if and how ZEBRAs form and 
what is their evolution and therefore it is too early to decide if their thickness can be reduced by advection.

\section{Discussion}
\label{sec:discussion}

Since the inner ($R \lesssim 30M$) regions of super-Eddington accretion flows onto black holes are likely to always be advection-dominated
\citep[cf][]{SLAN}, the very narrow funnels, that are supposed to produce strong radiation beaming near the black hole, most probably do not
exist in actually observed accreting systems. Nevertheless such strong beaming might be necessary to explain X-ray sources such as \object{SS~433} \citep{BKP06} and
ULXs \citep{K09,KL14,LKD15} if they are radiating at super-Eddington luminosities. Strong radiation beaming is certainly necessary to explain the
confirmed super-Eddington luminosity of the neutron-star \object{ULX-2 in the galaxy M82} \citep{KlL15,KL16}. It might also be the case of the ultraluminous supersoft source 
\object{ULS-1 in M81} \citep{LiuULS1}.
However, as mentioned above, we do not know how radiation
is emitted from very ($\dot m \gtrsim 30$) super-Eddington accretion flows, since in this case the emitting surface is located outside the computational grid.
One can only speculate that the beaming is determined by the vertical size of the outflow at the so-called \textsl{spherization radius} $R\sim 7\dot m M$ \citep{SS73}, in
other words by the ``walls'' formed by the outflows, as suggested by \citet{K08}.

\section{Conclusions}
\label{sec:summ}

Since the seminal \citet{SS73} paper various approaches have been used
to describe accretion flows for which the geometrical thinness ($H/R
\ll 1$) cannot be assumed. With respect to geometrical thickness the
resulting models could be divided into two classes: ``slim''
($H/R\lesssim 1$; this class also includes optically thin ADAFs) and
``thick'' ($H/R >1$) but the relation between the two was not
clear. It was implicitly assumed that by increasing the accretion rate
slim discs will inflate to become thick. However, this could not be
right because the thickness of slim discs is independent of the
accretion rate. 
Slim discs never become thick. In obtaining
this conclusion one assumes that accretion flows are stationary and
driven by local viscosity. It might not apply to flows forming in TDEs
\citep[see e.g.,][]{ZEBRA,Shio15} or to some flows that are dominated
by large-scale magnetic fields such as ion-tori of
\citet[][]{Reesetal82}\footnote{MAD configurations, however, compress accretion flows making them thin \cite[see e.g.][]{Tch+MAD,mckinney+MAD}.}. 
But it seems that standard accretion discs are
never obese. Whatever the accretion rate.

\begin{acknowledgements}
We are grateful to W{\l}odek Klu\'zniak for helpful comments.
We thank Maciek Wielgus for his help with PDs and Fig. \ref{fig:Wielgus}.
The anonymous referee's report was very helpful.
This research was supported by the Polish NCN grants UMO-2013/08/A/ST9/00795 and DEC-2012/04/A/ST9/00083.
JPL was supported in part by a grant from the French Space Agency CNES.
RSSV was supported by Funda\c{c}\~ao de Amparo \`a Pesquisa do Estado de São Paulo (FAPESP),
grants 2010/00487-9, 2013/01001-0 and 2015/10577-9.
RSSV is grateful for the hospitality at Harvard-Smithsonian Center for Astrophysics and at  Nicolaus Copernicus Astronomical Center. 
AS acknowledges support
for this work by NASA through Einstein Postdoctoral Fellowship number PF4-150126 awarded by the Chandra X-ray Center, which is operated by the
Smithsonian Astrophysical Observatory for NASA under contract
NAS8-03060. AS acknowledges computational support from NSF via XSEDE resources
(grant TG-AST080026N), and
from NASA via the High-End Computing (HEC) Program
through the NASA Advanced Supercomputing (NAS) Division at Ames
Research Center.
\end{acknowledgements}


\begin{thebibliography}{}


\bibitem[Abolmasov 
\& Chashkina(2015)] {Abolmasov15} Abolmasov, P., \& Chashkina, A.\ 2015, \mnras, 454, 3432

\bibitem[Abramowicz(1981)]{MA81} Abramowicz, M.~A.\ 1981, 
Nature, 294, 235 

\bibitem[Abramowicz(1985)]{MA85} Abramowicz, M.~A.\ 1985, 
PASJ, 37, 727 

\bibitem[Abramowicz(2005)]{MA05} Abramowicz, M.~A.\ 2005, in 
Growing Black Holes: Accretion in a Cosmological Context, ed. A. Merloni, S. Nayakshin, R. A, ESO astrophysics symposia. Springer (Berlin, Germany), 257

\bibitem[Abramowicz 
\& Fragile(2013)]{AF} Abramowicz, M.~A., \& Fragile, P.~C.\ 2013, Living Reviews in Relativity, 16, 1 

\bibitem[Abramowicz 
\& Piran(1980)]{AP80} Abramowicz, M.~A., \& Piran, T.\ 1980, \apjl, 241, L7 

\bibitem[Abramowicz et 
al.(1978)]{AJS} Abramowicz, M., Jaroszy\'nski, M., \& Sikora, M.\ 1978, \aap, 63, 221 

\bibitem[Abramowicz et al.(1983)]{AHG83} Abramowicz, M.~A., 
Henderson, P.~F., \& Ghosh, P.\ 1983, \mnras, 203, 323 

\bibitem[Abramowicz et al.(1997)]{ALP} Abramowicz, M.~A., 
Lanza, A., \& Percival, M.~J.\ 1997, \apj, 479, 179 

\bibitem[Abramowicz et al.(2000)]{adiosM} Abramowicz, M.~A., 
Lasota, J.-P., \& Igumenshchev, I.~V.\ 2000, \mnras, 314, 775 

\bibitem[Abramowicz et al.(1988)]{slim1} Abramowicz, M.~A., 
Czerny, B., Lasota, J.~P., \& Szuszkiewicz, E.\ 1988, ApJ, 332, 646 

\bibitem[Abramowicz et al.(1995)]{adaf1} Abramowicz, M.~A., 
Chen, X., Kato, S., Lasota, J.-P., \& Regev, O.\ 1995, ApJ, 438, L37 

\bibitem[Abramowicz et 
al.(2010)]{isco} Abramowicz, M.~A., Jaroszy{\'n}ski, M., Kato, S., et al.\ 2010, A\&A, 521, A15 

\bibitem[Begelman(1979)]{Beg79} Begelman, M.~C.\ 1979, 
\mnras, 187, 237

\bibitem[Begelman et al.(2006)]{BKP06} Begelman, M.~C., King, 
A.~R., \& Pringle, J.~E.\ 2006, \mnras, 370, 399 

\bibitem[Chen et al.(1995)]{Chenetal95} Chen, X., Abramowicz, 
M.~A., Lasota, J.-P., Narayan, R., \& Yi, I.\ 1995, \apjl, 443, L61 

\bibitem[Coughlin 
\& Begelman(2014)]{ZEBRA} Coughlin, E.~R., \& Begelman, M.~C.\ 2014, \apj, 781, 82 

\bibitem[De Villiers 
\& Hawley(2003)]{DVH} De Villiers, J.-P., \& Hawley, J.~F.\ 2003, \apj, 592, 1060 

\bibitem[Dotan 
\& Shaviv(2011)]{DS11} Dotan, C., \& Shaviv, N.~J.\ 2011, \mnras, 413, 1623 

\bibitem[Ichimaru(1977)]{I77} Ichimaru, S.\ 1977, \apj, 
214, 840 

\bibitem[Fishbone 
\& Moncrief(1976)]{fishbone} Fishbone, L.~G., \& Moncrief, V.\ 1976, \apj, 207, 962 

\bibitem[Fragile et al.(2014)]{fragile+14} Fragile, P.~C., Olejar, 
A., \& Anninos, P.\ 2014, arXiv:1408.4460 

\bibitem[Frank et al.(2002)]{FKR02} Frank, J., King, A., 
\& Raine, D.~J.\ 2002, Accretion Power in Astrophysics, Cambridge University Press (Cambridge, UK)

\bibitem[Gu(2012)]{Gu12} Gu, W.-M.\ 2012, \apj, 753, 118

\bibitem[Gu \& Lu (2007)]{Gu07}
Gu, W.-~M., Lu, J.-~F. \ 2007, ApJ, 660, 541 

\bibitem[Hawley(2000)]{Hawley2000} Hawley, J.~F.\ 2000, \apj, 528, 
462 

\bibitem[Jaroszy\'nski et al.(1980)]{JAP} Jaroszy\'nski, M., 
Abramowicz, M.~A., \& Paczy\'nski, B.\ 1980, \actaa, 30, 1 

\bibitem[Kato et al.(2008)]{Katoetal} Kato, S., Fukue, J., 
\& Mineshige, S.\ 2008, Black-Hole Accretion discs -- Towards a New Paradigm, Kyoto University Press (Kyoto, Japan) 

\bibitem[Kawaguchi(2003)]{kawa03} Kawaguchi, T.\ 2003, ApJ, 
593, 69 

\bibitem[King(2008)]{K08} King, A.\ 2008, \nar, 51, 775 

\bibitem[King(2009)]{K09} King, A.~R.\ 2009, \mnras, 393, 
L41

\bibitem[King 
\& Lasota(2014)]{KL14} King, A., \& Lasota, J.-P.\ 2014, \mnras, 444, L30 

\bibitem[King 
\& Lasota(2016)]{KL16} King, A., \& Lasota, J.-P.\ 2016, \mnras, submitted 

\bibitem[Klu{\'z}niak 
\& Lasota(2015)]{KlL15} Klu{\'z}niak, W., \& Lasota, J.-P.\ 2015, \mnras, 448, L43 

\bibitem[Komissarov(2006)]{K06} Komissarov, S.~S.\ 2006, \mnras, 368, 993 

\bibitem[Lasota(2015)] {L15}Lasota, J.-P.\ 2015, Black Accretion Discs, in Astrophysical Black Holes -- From fundamental aspects
to latest developments, ed. C. Bambi (Springer, Astrophysics and Space Science Library), in press, arXiv:1505.02172

\bibitem[Lasota \& al. (2015)]{LKD15}
Lasota J.-P., King A. R., Dubus G., 2015, \apjl, 801, L4

\bibitem[Liu et al.(2015)]{LiuULS1} Liu, J.-F., Bai, Y., Wang, 
S., et al.\ 2015, \nat, 528, 108 

\bibitem[Lynden-Bell \& Pringle(1974)]{LBP74} Lynden-Bell, D., \& Pringle, J.~E.\ 1974, \mnras, 168, 603 

\bibitem[McKinney et al.(2012)]{mckinney+MAD} McKinney, J.~C., 
Tchekhovskoy, A., \& Blandford, R.~D.\ 2012, \mnras, 423, 3083 

\bibitem[McKinney et al.(2014)]{mckinney+harmrad} McKinney, J.~C., 
Tchekhovskoy, A., S{\c a}dowski, A., \& Narayan, R. \ 2014, $\rm \mnras$, 441, 3177 

\bibitem[Narayan \& Yi(1994)]{ny0} Narayan, R., \& Yi, I.\ 1994, \apjl, 428, L13 

\bibitem[Narayan \& Yi(1995)]{ny1} Narayan, R., \& Yi, I.\ 1995, ApJ, 452, 710 

\bibitem[Novikov 
\& Thorne(1973)]{NT73} Novikov, I.~D., \& Thorne, K.~S.\ 1973, in Black Holes (Les Astres Occlus), ed. C. Dewitt, B.~S. Dewitt,  Gordon and Breach (New York, USA), 343

\bibitem[Ohsuga et al.(2005)]{Ohsuga+05} Ohsuga, K., Mori, M., 
Nakamoto, T., \& Mineshige, S.\ 2005, \apj, 628, 368 

\bibitem[Paczy\'nski(1998)]{BP} Paczy{\'n}ski, B.\ 1998, Acta Astronomica, 
48, 667 (\emph{use the astro-ph/9812047 version})

\bibitem[Paczy{\'n}ski 
\& Wiita(1980)]{PW} Paczy{\'n}ski, B., \& Wiita, P.~J.\ 1980, \aap, 88, 23 

\bibitem[Paczy\'nski 
\& Bisnovatyi-Kogan(1981)]{PBK81} Paczy\'nski, B., \& Bisnovatyi-Kogan, G.\ 1981, \actaa, 31, 283 

\bibitem[Qian et 
al.(2009)]{Qianetal09} Qian, L., Abramowicz, M.~A., Fragile, P.~C., et al.\ 2009, \aap, 498, 471

\bibitem[Rees et al.(1982)]{Reesetal82} Rees, M.~J., Begelman, 
M.~C., Blandford, R.~D., \& Phinney, E.~S.\ 1982, \nat, 295, 17 

\bibitem[Rezzolla 
\& Zanotti(2013)]{RZ13} Rezzolla, L., \& Zanotti, O.\ 2013, Relativistic Hydrodynamics, Oxford University Press (Oxford, UK)

\bibitem[S{\c a}dowski(2009)]{olek09} S{\c a}dowski, A.\ 2009, 
ApJS, 183, 171 

\bibitem[S{\c a}dowski(2011)]{olekthesis} S{\c a}dowski, A.\ 2011, 
Slim discs around Black Holes, PhD thesis, arXiv:1108.0396  

\bibitem[S{\c a}dowski et 
al.(2011)]{olek11} S{\c a}dowski, A., Abramowicz, M., Bursa, M., et al.\ 2011, A\&A, 527, A17 

\bibitem[S{\c a}dowski et al.(2015a)]{olek14} S{\c a}dowski, 
A., Narayan, R., Tchekhovskoy, A., et al.\ 2015a, \mnras, 447, 49 

\bibitem[S{\c a}dowski 
\& Narayan(2015b)]{sadowski+radjets} S{\c a}dowski, A., \& Narayan, R.\ 2015b, \mnras, 453, 3213 

\bibitem[Sadowski 
\& Narayan(2015c)]{sadowski+3d} Sadowski, A., \& Narayan, R.\ 2015c, arXiv:1509.03168 

\bibitem[Sadowski et al.(2016)]{SLAN} Sadowski, A., Lasota, 
J.-P., Abramowicz, M.~A., \& Narayan, R.\ 2016, \mnras, in press, arXiv:1510.08845 

\bibitem[Shakura 
\& Sunyaev(1973)]{SS73} Shakura, N.~I., \& Sunyaev, R.~A.\ 1973, \aap, 24, 337 

\bibitem[Shiokawa et al.(2015)]{Shio15} Shiokawa, H., Krolik, 
J.~H., Cheng, R.~M., Piran, T., \& Noble, S.~C.\ 2015, \apj, 804, 85 

\bibitem[Sikora(1981)]{S81} Sikora, M.\ 1981, \mnras, 196, 257

\bibitem[Tchekhovskoy et al.(2011)]{Tch+MAD} Tchekhovskoy, A., 
Narayan, R., \& McKinney, J.~C.\ 2011, \mnras, 418, L79 


\bibitem[Wielgus et al(2015)]{Wetal15}Wielgus, M., Yan, W., Lasota J.-P.  \& Abramowicz M.A., 2015,
A\&A, submitted, arXiv:1512.00749
 
\end{thebibliography}
\end{document}